\documentclass[prl,superscriptaddress,twocolumn,subeqn,amssymb,amsmath,amsfonts,aps]{revtex4}

\usepackage[pdftex]{graphicx} % Include figure files
\usepackage{epsfig}
\usepackage{dcolumn}  % Align table columns on decimal point
\usepackage{longtable}% Allow column and page-spanning tables.
\usepackage{bm}       % bold math
\usepackage{subfigure}% subfigures (labels them (a),(b),...)
\usepackage{multirow} %
\usepackage{epstopdf}

\usepackage{bm}       % bold math
\usepackage{subfigure}% subfigures (labels them (a),(b),...)
\usepackage{float}

\newcommand\captionof[1]{\def\@captype{#1}\caption}

\newcommand{\cmangle}{\cos \theta^\phi_{\mbox{\scriptsize c.m.}}}

\newcommand{\sqrts}{\sqrt{s}}
\newcommand{\kkb}{K\overline{K}}

\newcommand*{\NOWSLAC}{SLAC National Accelerator Laboratory, Menlo Park, CA 94025}

\begin{document}
%
%\preprint{CMU/JLab/??????}
%
%\input{authors}...eventually get this for CLAS
%
% ...but for now...
%
%\input{author_list.tex}
\author{Biplab~Dey} 
%\affiliation{SLAC National Accelerator Laboratory, Menlo Park, CA 94025}
\affiliation{Carnegie Mellon University, Pittsburgh, PA 15213}
\altaffiliation[Current address: ]{\NOWSLAC}
\email{biplabd@slac.stanford.edu}
%\author{C.~A.~Meyer} 
%\affiliation{Carnegie Mellon University, Pittsburgh, PA 15213}
%\author{M.~Bellis} 
%\affiliation{Carnegie Mellon University, Pittsburgh, PA 15213}
%\author{M.~Williams} 
%\affiliation{Carnegie Mellon University, Pittsburgh, PA 15213}
\date{\today}

\title{Phenomenology of $\phi$ photoproduction from recent CLAS data at Jefferson Lab}
%%%%%%%%%%%%%%%%%%%%%%%%%%%%%%%%%%%%%%%%%%%%%%%%%%%%%%%%%%%%%%%%%%%%%%%%%%%%%%%
%
% abstract
%
%%%%%%%%%%%%%%%%%%%%%%%%%%%%%%%%%%%%%%%%%%%%%%%%%%%%%%%%%%%%%%%%%%%%%%%%%%%%%%%
\begin{abstract} 
We comment on the important phenomenological aspects of the recent high-statistics and wide-angle coverage $\phi$ photoproduction data from CLAS at Jefferson Lab. The most prominent feature is a localized structure at a center-of-mass (c.m) energy $\sqrts \sim 2.2$~GeV that is not expected in a simple $t$-channel Pomeron-exchange model. The structure exists only at the forward production angles that almost rules out any resonance contribution. Strong rescattering effects between the $p \phi$ and $K^+\Lambda(1520)$ channels could be possible explanations. The analyses of both charged- ($\phi \to K^+ K^-$) and neutral- ($\phi \to K^0_S K^0_L$) $\kkb$ decay modes of the $\phi$, that show some minor differences, can be illuminating in this respect. We also comment on the angular structure of the Pomeron-parton coupling as borne out in the polarization data where the often-asumed $s$-channel helicity conservation is seen to be broken.
\end{abstract} 
%
%\pacs{
%      {????????}
%      {????????}
%     } % end of PACS codes
\maketitle
%%%%%%%%%%%%%%%%%%%%%%%%%%%%%%%%%%%%%%%%%%%%%%%%%%%%%%%%%%%%%%%%%%%%%%%%%%%%%%%
%\input{intro}
\begin{figure}
\begin{center}
\subfigure[]{
{\includegraphics[width=2.5in,angle=90]{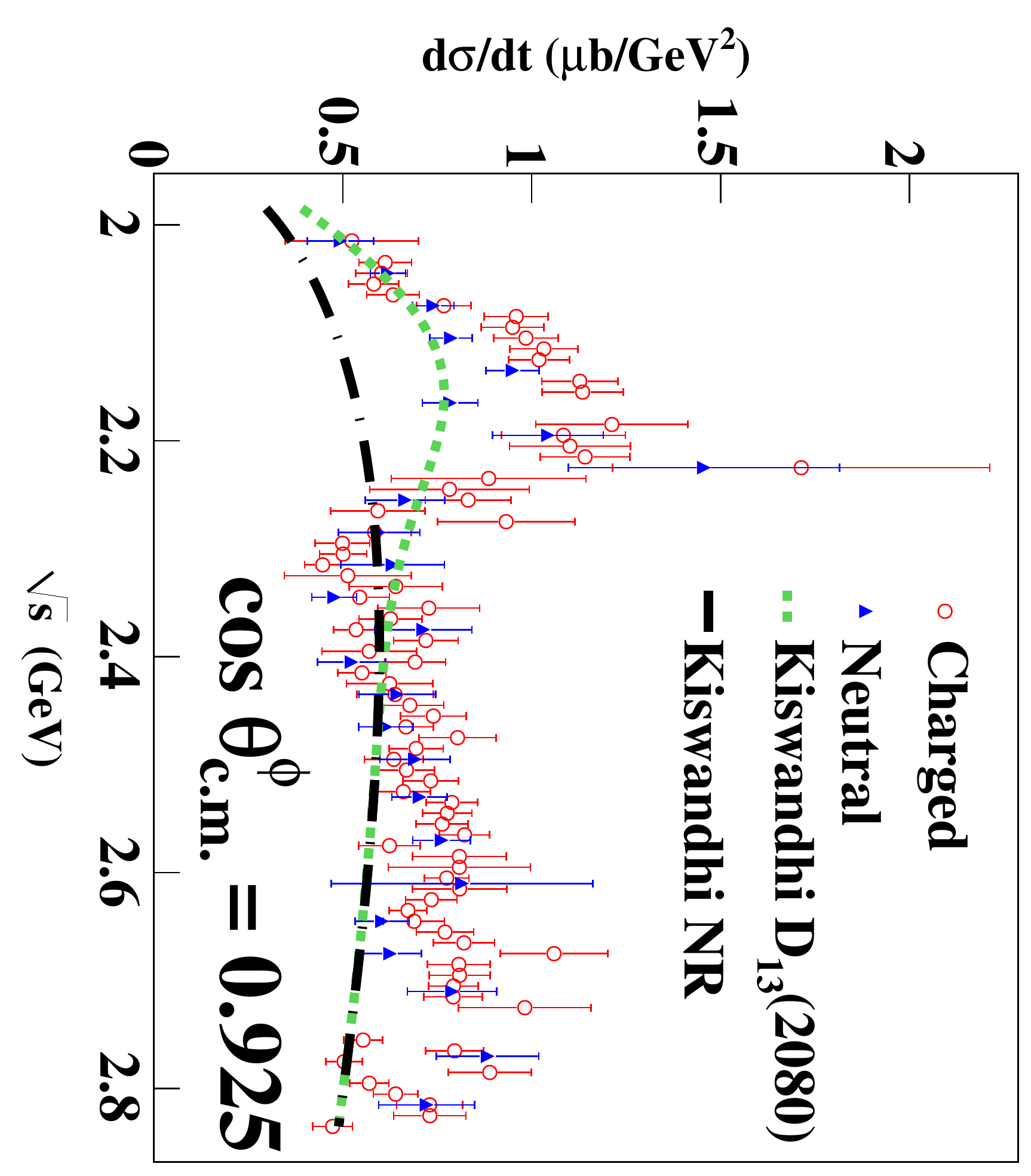}}
%{\includegraphics[width=2.5in,angle=90]{dsigdt_fwd_angle_clas_kiswandhi_paper.eps}}
}
\subfigure[]{
{\includegraphics[width=2.5in,angle=90]{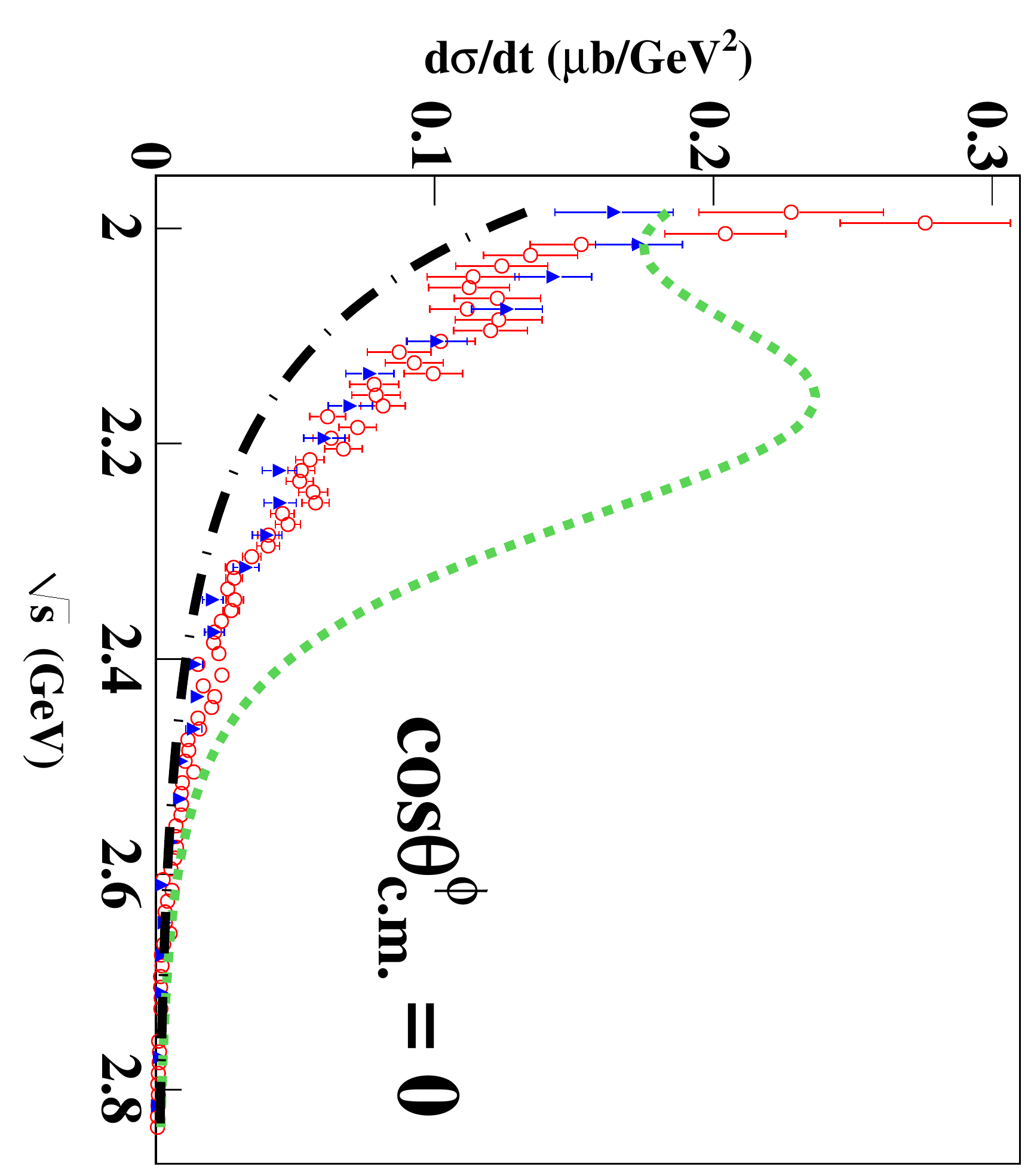}}
%{\includegraphics[width=2.5in,angle=90]{dsigdt_mid_angle_clas_kiswandhi_paper.eps}}
}
\caption{\label{fig:peak}(Color online) Comparison between the charged- and neutral-mode $\phi$ data from CLAS~\cite{phi_prc} and model predictions from Kiswandhi {\em et al.}~\cite{kiswandhi_prc} that include a $D_{13}(2080)$ resonance exchange in the $s$-channel. The data shows a local structure at (a) forward-angles but none at (b) mid-angles.}
\end{center}
\end{figure}

Photo- and electroproduction of vector mesons have long been an important laboratory for exploring QCD via hadron dynamics. The photons (real or virtual) behave like a beam of vector mesons $V\in \{\rho,\omega,\phi,J/\psi...\}$ and process is akin to vector meson scattering. Among the light mesons, the $\phi$ is especially attractive because of its almost pure $|s\bar{s}\rangle$ valence quark content. Assuming a suppressed strange-quark content in the proton, $\phi p$ scattering is OZI suppressed. The dominant $t$-channel exchange is via gluon-rich objects like the so-called universal Pomeron trajectory, that is expected to dominate $\phi$ photoproduction at all energies.

The same OZI-suppression also means that the $\phi$ cross-sections are much lower than those for $\rho$ or $\omega$. As a result, world data on the $\phi$ have been extremely scarce. The recent high-statistics and wide kinematic coverage results from CLAS~\cite{phi_prc} have completely altered this situation. With detailed results on both cross-sections and the spin density matrix elements (SDME), $\phi$ photoproduction is now as much as as a precision measurement as for the OZI-favored $\rho$ and $\omega$. In this Letter, we comment on some of the important features in these new results.

The most prominent feature is a local structure at around the center-of-mass (c.m) energy $\sqrts \sim 2.2$~GeV that is not expected in a simple $t$-channel Pomeron-exchange model. Similar indications were also seen in previous results from LEPS~\cite{mibe}. Two different explanations have been offered in the literature. Kiswandhi {\em et al.}~\cite{kiswandhi_prc} has ascribed this to $s$-channel exchange of a high strangeness content $N^\ast$ resonance, while Ozaki {\em et al.}~\cite{ozaki} and Ryu {\em et al.}~\cite{ryu} have conjectured this as a rescattering between $\phi p$ and $K^+ \Lambda(1520)$  with the same $p K^+ K^-$ final-state configuration.

The CLAS results both confirm and expand on the study of this feature in several ways. First, the LEPS data~\cite{mibe} pertained only to the forward most angular bin in the $\phi$ scattering angle $\cmangle$. Fitting only to the LEPS data, the Kiswandhi model~\cite{kiswandhi_prc} interpreted the structure as due to a spin-3/2 resonance. However, once the wide-angle CLAS data is incorporated, the structure is found to be non-existent away from the forward-angle region, a feature that is very difficult to explain within the context of a resonance interpretation. A single resonance with a given spin-parity will almost never produce a structure only at forward angles. This is borne out in going from Fig.~\ref{fig:peak}a to Fig.~\ref{fig:peak}b, where the resonanant structure from the Kiswandhi model (dashed-green) is present at all angles.

Second, availability of the results from neutral decay mode of the $\phi$ into $K^0_S K^0_L$ from CLAS provides valueable insight. Unlike the charged mode, the neutral mode does not involve a common final state configuration with the $\Lambda(1520)\to p K^-$. If the rescattering hypothesis~\cite{ozaki,ryu} between the $\phi p$ and $K^+ \Lambda(1520)$ is correct, we expect differences between the charged ($\phi \to K^+ K^-$) and neutral ($\phi \to K^0_S K^0_L$) differential cross-sections. While some mild charged/neutral differences are seen, they are not large enough to explain the presence of the local structure solely due to rescattering. Moreover, the $\sqrts \sim 2.2$~GeV structure is prominently seen in the neutral-mode results as well, which should be unaffected by interference effects. Therefore, some other mechinsm must contribute as well. The CLAS analysis also studied the effect of a hard cut on $|M(pK^-) -1.52|> 0.015$~GeV that removed the region of kinematic overlap between the charged mode and $\Lambda(1520)$ channels. No discernably large effect due to the hard cut was found. 
\begin{figure}
\begin{center}
\includegraphics[width=3.3in]{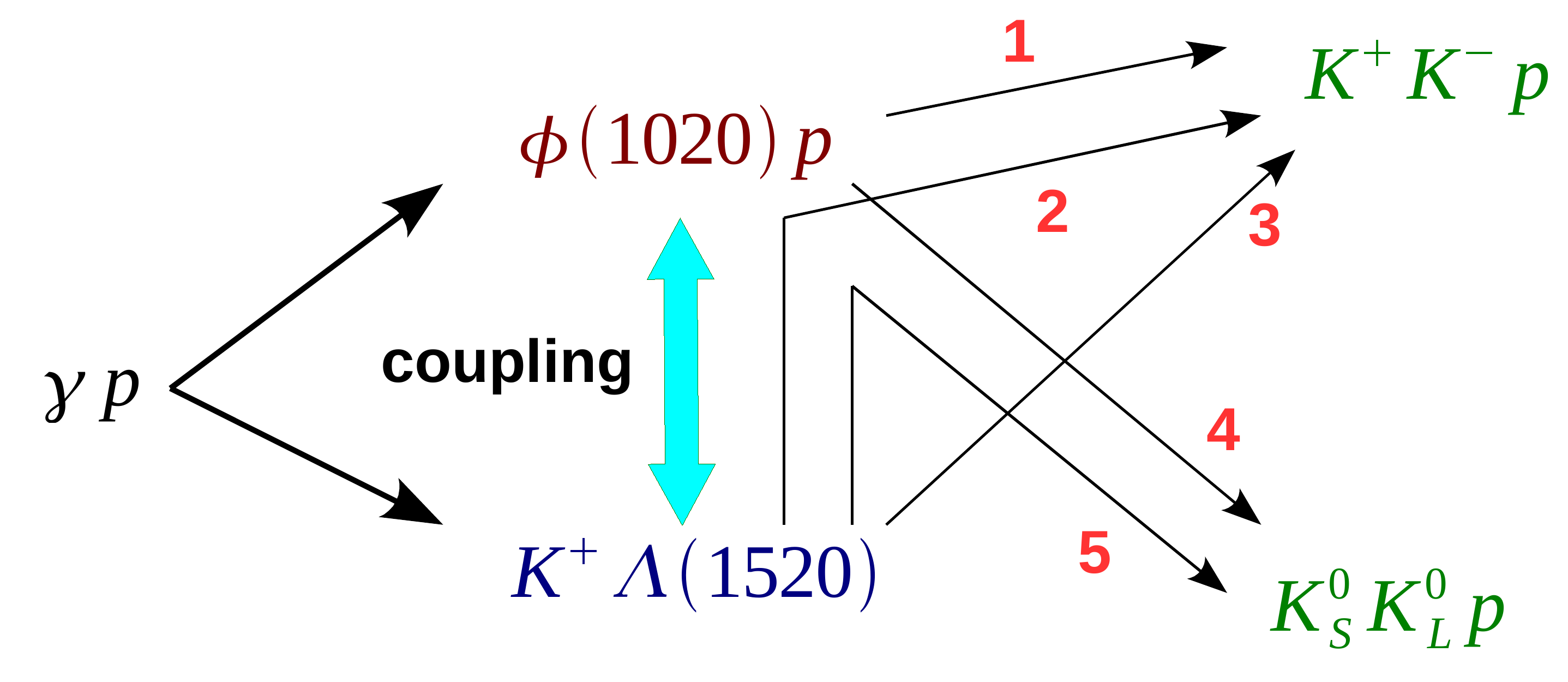}
\caption{\label{fig:phi_L1520_rescattering}(Color online) Pictorial representation of the different ``paths'' to the $\kkb p$ final states and possible coupling between the $\phi p$ and $K^+ \Lambda(1520)$ channels.}
\end{center}
\end{figure}
Fig.~\ref{fig:phi_L1520_rescattering} shows a pictorial representation of the possible channel-coupling between $\phi p$ and $K^+ \Lambda(1520)$. The process $\gamma p \to K^+ \Lambda(1520) \to K^0_S K^0_L p$ is possible, if there is appreciable rescattering between the two channels.

Interestingly, one can also look at the different decays modes of the $\Lambda(1520)$ into $p K^-$ or $\Sigma \pi$. If there is a channel coupling between the $\phi p$ and the $K^+ \Lambda(1520)$, the different $\Lambda(1520)$ decay modes should show different results. CLAS has recently published results on the $\Sigma \pi$ modes~\cite{kei_l1520}, while the $p K^-$ analysis is under progress. A local enhancement of the cross-section at $\sqrts \sim 2.2$~GeV is also seen in the $K^+ \Lambda(1520)$. Further, there are hints that the $\Sigma \pi$ and $p K^-$ modes show differences around $\sqrts \sim 2.2$~GeV (see Fig.~7.14 in Ref.~\cite{kei-thesis} in the $E_\gamma \in [1.994,2.229]$~GeV bin). Our surmise is that the subtle differences between the different decay modes of the $\phi$ and $\Lambda(1520)$ owe their origin to the same underlying mechansim. A full coupled channel analysis between the $\phi p$ and $K^+ \Lambda(1520)$ incorporating the different modes should be very interesting.

Kiswandhi {\em et al.}~\cite{kiswandhi_prc} attempted to explain the structure as being due a $N^\ast$ resonance with large strangeness content. They consider the $J^P = \frac{3}{2}^\pm$ states with masses around $\sqrts = 2080$~MeV fit to the forward-angle data from LEPS. Fig.~\ref{fig:peak} shows the model results in comparison with the latest CLAS results. The red empty circles and the blue filled triangles are the charged- and neutral-mode data respectively. The black dot-dashed curves are the non-resonant (NR) $t$-channel contribution, primarily from the Pomeron, while the green dotted curve includes a $D_{13}(2080)$ resonance. At forward-angles (Fig.~\ref{fig:peak}a), both the data and the model-curves show a peaking structure. However, away from the forward-angle region (Fig.~\ref{fig:peak}b), the data is smooth aross the $\sqrts \sim 2.1$~GeV region, while the resonance structure bears out prominently in the model calculations. We note again that the model fits were performed only to the LEPS forward-angle data. The availablilty of the new wide-angle data from CLAS strongly indicate against an $s$-channel resoance, since for simple spin-parity states or their combinations, it is quite unlikely to produce a feature that persists only in the forward angles. In addition, a nucleon resonance that couples so strongly to the $\phi$ must have a very large strangeness content. Although not totally impossible, this is certainly unnatural.

We are therefore left with an interesting puzzle here. If the structure is not due to a resonance, $\phi p$-$K^+\Lambda(1520)$ rescattering could be a plausible scenario, where the rescattering process occurs blind to how the $\phi$ or $\Lambda(1520)$ states decay subsequently. This could explain the structure in the neutral mode as well, because ab initio, the neutral mode does not have any connection with the $\Lambda(1520)$.

\begin{figure}
\begin{center}
\includegraphics[width=2.6in,angle=90]{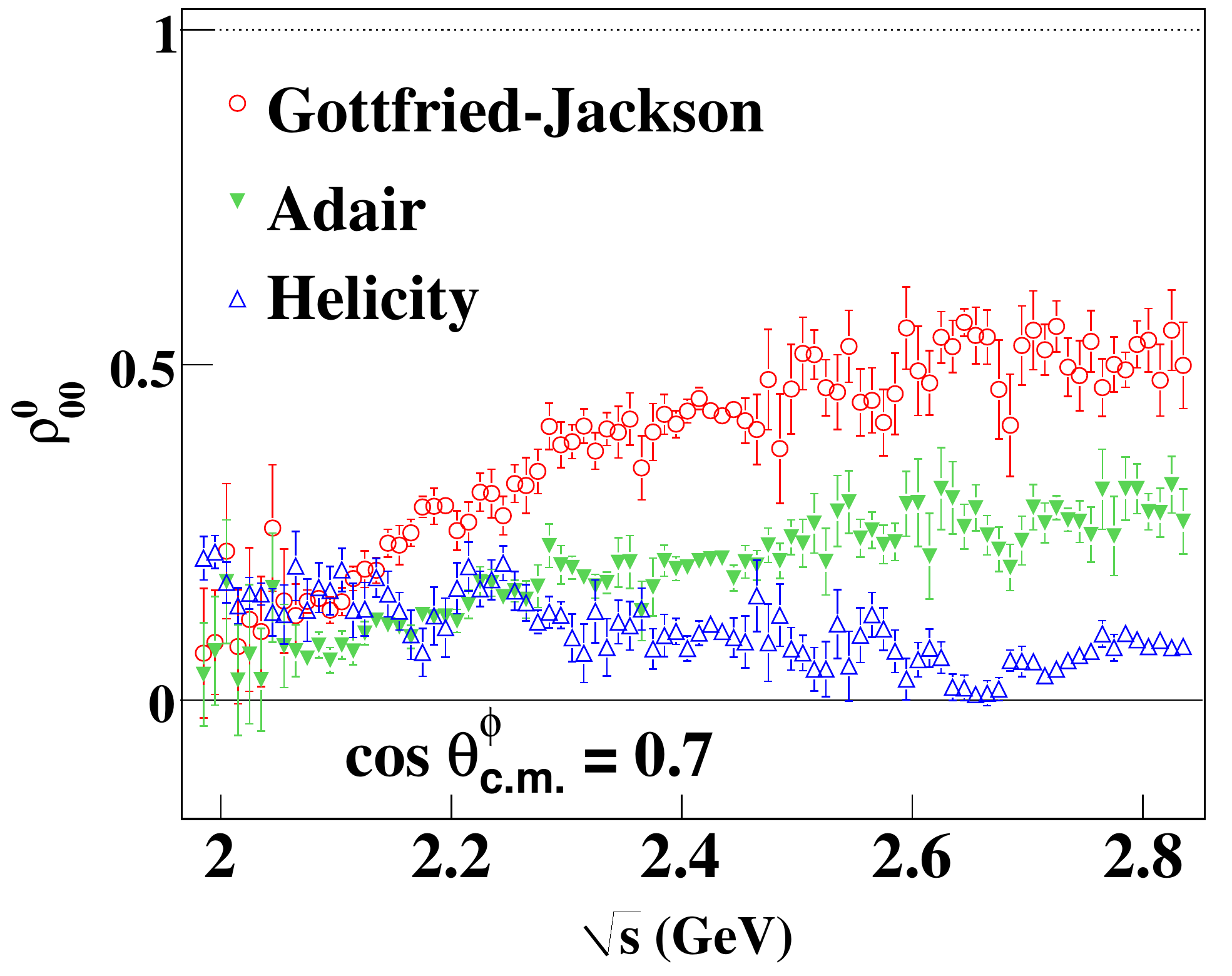}
\caption{\label{fig:hel_con}(Color online) Since $\rho^0_{00} \neq 0$ in the diffraction regime, helicity-conservation is broken in all three reference frame.}
\end{center}
\end{figure}

Above $\sqrts \sim 2.3$~GeV, $d\sigma/dt$ shows only a very slow rise with energy that is typical of diffractive phenomenology via Pomeron exchange. Conventionally, the Pomeron is understood as a gluon-rich Regge trajectory with quantum numbers of the vacuum ($J^{PC} = 0^{++}$). Exchange of a spin-0 object in the $t$-channel should lead to zero helicity flip between the incoming ($\gamma$) and outgoing ($\phi$) vector mesons. However, experimentally, $t$-channel helicity conservation (TCHC) is long known to be strongly broken and one sees instead $s$-channel helicity conservation (SCHC). It is somewhat unexpected that a $t$-channel process should conserve helicity in the $s$-channel. The helicity-flip and non-flip amplitudes for the Pomeron must be related in a special manner to facilitate SCHC, as shown by Gilman {\em et al.}~\cite{gilman}. Helcity conservation can be studied by looking at the spin density matrix element (SDME) $\rho^0_{00}$ that is proportional to the sum of the squares of the helicity-flip amplitudes. The SDME's are not Lorentz invariant quantities and choice of the spin-quantization axis depends on the production mechanism one is studying. In the Adair frame, the spin-quantization axis is along the incoming photon beam direction, while for the Helicity frame, it is the direction of the outgoing $\phi$ in the c.m. frame. The Gottfried-Jackson frame quantization axis corresponds to the incoming photon direction as seen in the $\phi$ rest frame. SCHC and TCHC corresponds to $\rho^0_{00} =0$ in the Helicity and Gottfried-Jackson frames, respectively. The CLAS data confirms that TCHC is badly broken, but also shown that even in the diffractive region (large $\sqrts$ and forward angles) the usual assumption of SCHC in vector meson photoproduction is not valid. Fig.~\ref{fig:hel_con} shows this in the $\cmangle = 0.7$ bin, where $\rho^0_{00} \neq 0$ in all three reference frames.

There is no fundamental reason to expect helicity conservation in any of the three frames since this depends on how the Pomeron couples to partons. Violation of TCHC confirms that the coupling is not a scalar, but new data also shows that the often-taken assumption of SCHC in vector meson photoproduction is also invalid.

To summarize, in this Letter, we reported on some unique features in the new $\phi$ photoproduction results from CLAS, where hints of a strong coupling between two different channels are clearly exhibited. While previous world data showed some of these features as well, the wide-angle coverage almost completely rules out a resonance interpretation that was a strong contender as a potential explanation, based on the forward-angle data only. The new polarization data will also shed light on how the Pomeron couples to partons.

\begin{acknowledgments}
The author thanks Alvin Kiswandhi for providing the theory model curves.
\end{acknowledgments}

\end{document}